\documentclass[twocolumn,showpacs,preprintnumbers,amsmath,amssymb]{revtex4}

\usepackage{graphicx}
\usepackage{dcolumn}
\usepackage{bm}

\newcommand{\bq}{\begin{equation}}
\newcommand{\eq}{\end{equation}}
\newcommand{\bqa}{\begin{eqnarray}}
\newcommand{\eqa}{\end{eqnarray}}
\newcommand{\nn}{\nonumber \\}

\def\be     {\begin{equation}}
\def\ee     {\end{equation}}
\def\bea        {\begin{eqnarray}}
\def\eea        {\end{eqnarray}}
\def\bnn    {\begin{eqnarray*}}
\def\enn    {\end{eqnarray*}}

\begin{document}

\title{Existence of Fermion Zero Modes and Deconfinement of Spinons in Quantum Antiferromagnetism resulting from Algebraic Spin Liquid}
\author{Ki-Seok Kim}
\affiliation{School of Physics, Korea Institute for Advanced
Study, Seoul 130-012, Korea}
\date{\today}

\begin{abstract}
We investigate the quantum antiferromagnetism arising from
algebraic spin liquid via spontaneous chiral symmetry breaking. We
claim that in the antiferromagnet massive Dirac spinons can appear
to make broad continuum spectrum at high energies in inelastic
neutron scattering. The mechanism of spinon deconfinement results
from the existence of fermion zero modes in single monopole
potentials. Neel vectors can make a skyrmion configuration around
a magnetic monopole of compact U(1) gauge fields. Remarkably, in
the monopole-skyrmion composite potential the Dirac fermion is
shown to have a zero mode. The emergence of the fermion zero mode
forbids the condensation of monopoles, resulting in deconfinement
of Dirac spinons in the quantum antiferromagnet.
\end{abstract}


\maketitle

\section{Introduction}

High $T_c$ superconductivity is believed to result from hole
doping to an antiferromagnetic Mott insulator. Hole doping to an
antiferromagnetic Mott insulator destroys antiferromagnetic long
range order, resulting in one quantum disordered paramagnetic Mott
insulator that is considered to be the pseudogap phase in high
$T_c$ cuprates. High $T_c$ superconductivity is expected to arise
from further hole doping to the paramagnetic Mott
insulator\cite{Review}. In this respect it should be a starting
point for the theory of high $T_c$ superconductivity to understand
the nature of the antiferromagnetic Mott insulator.

Recently, the paramagnetic Mott insulator was proposed to be {\it
algebraic spin liquid}, where spin $1$ antiferromagnetic
fluctuations {\it break up} into more elementary spin $1/2$
fractionalized excitations called {\it spinons}\cite{Review}. If
the algebraic spin liquid correctly describes spin degrees of
freedom in the pseudogap phase, its parent antiferromagnet may
include the trace of fractionalized spinon excitations. In the
present paper we find a trace of deconfined spinons in the quantum
antiferromagnet.

There are some experimental reports that the spin spectrum in the
antiferromagnetic Mott insulator is difficult to understand only
by antiferromagnons\cite{Experiment}. Although the dispersing
peaks observed in inelastic neutron scattering measurements can be
interpreted as antiferromagnons, an analysis of the spectral
weight shows that long range order and antiferromagnons can
account for only about $1/2$ of the observed
spectrum\cite{Experiment}. This difficulty originates from the
presence of finite spectral weight at high
energies\cite{Experiment}. The unidentified spectral weight
indicates the presence of excitations beyond one magnon
mode\cite{Experiment}.

In the present paper we claim that {\it the unidentified spectral
weight at high energies in the spin spectrum can be identified
with deconfined gapped spinons}. In a theoretical point of view,
if antiferromagnetism is supposed to originate from algebraic spin
liquid via spontaneous chiral symmetry breaking, the emergence of
deconfined spinons seems to be possible. {\it The mechanism of
spinon deconfinement results from the existence of fermion zero
modes in single monopole potentials}. It should be noted that the
mechanism of monopole suppression in the present antiferromagnet
completely differs from that in the algebraic spin liquid, where
existence of quantum criticality (critical fluctuations of matter
fields) is the origin of spinon deconfinement\cite{Hermele_QED3}.
In high energy physics the mechanism of monopole suppression by
fermion zero modes is well
known\cite{Jackiw,tHooft,Callias,Witten}. In the context of
quantum antiferromagnetism there was a try to find fermion zero
modes. Marston has tried to find fermion zero modes {\it in the
algebraic spin liquid}\cite{Marston}. Unfortunately, the fermion
zero mode was not found.

In the present communication {\it we find a fermion zero mode in
the antiferromagnetism arising from the algebraic spin liquid via
spontaneous chiral symmetry breaking}. In the present
antiferromagnet the effective action is pretty much similar to a
well studied action in high energy physics, where a fermion zero
mode exists. In the antiferromagnet Neel vectors can make a
hedgehog configuration around a magnetic monopole of compact U(1)
gauge fields. Remarkably, in the monopole-skyrmion composite
potential the Dirac fermion is shown to have a zero mode. The
emergence of the fermion zero mode forbids condensation of
magnetic monopoles, resulting in deconfinement of Dirac spinons.
Notice that Dirac spinons are massive owing to spontaneous chiral
symmetry breaking. These gapped spinons would appear to make broad
continuum spectrum above their mass gap in inelastic neutron
scattering\cite{Experiment,Fisher,YBKim}.

The main body of the present paper consists of three major parts.
The first is to introduce a fermionic nonlinear $\sigma$ model
[Eq. (1)] as a low energy effective field theory for the quantum
antiferromagnetism. The second is to prove the existence of
fermion zero modes based on the proposed fermionic nonlinear
$\sigma$ model [Eq. (3)]. The last is to investigate the effect of
the fermion zero mode [Eq. (13)].

\section{Existence of Fermion zero modes and Deconfinement of spinons}

\subsection{Fermionic Nonlinear $\sigma$ Model for Quantum Antiferromagnetism}

We consider the following effective action called fermionic
nonlinear $\sigma$ model \bqa && Z = \int{D {\vec
n}}{D\lambda}{D\psi_n}{Da_{\mu}} e^{- S_{ASL} - S_{M} -
S_{NL\sigma{M}}} , \nn && S_{ASL} = \int{d^3x} \Bigl[
\bar{\psi}_{n}\gamma_{\mu}(\partial_{\mu} - ia_{\mu})\psi_{n} +
\frac{1}{2e^2}|\partial\times{a}|^2 \Bigr] , \nn && S_{M} =
\int{d^3x} m_{\psi}\bar{\psi}_{n}({\vec n}\cdot{\vec
\tau}_{nm})\psi_{m} , \nn && S_{NL\sigma{M}} = \int{d^3x} \Bigl[
\frac{Nm_{\psi}}{4\pi}|\partial_{\mu}{\vec n}|^{2} -
i\lambda(|{\vec n}|^{2} - 1) \Bigr] . \eqa In $S_{ASL}$ $\psi_{n}$
is a massless Dirac spinon with a flavor index $n = 1, ..., N$
associated with SU(N) spin symmetry. $a_{\mu}$ is a {\it compact}
U(1) gauge field mediating long range interactions between Dirac
spinons. $e$ is an {\it internal} electric charge of the Dirac
spinon. $S_{ASL}$ in Eq. (1) was proposed to be an effective field
theory for one possible quantum paramagnetism, called algebraic
spin liquid, of SU(N) quantum antiferromagnets on two dimensional
square lattices\cite{Review}. However, the stability of the
algebraic spin liquid has been suspected owing to instanton
excitations of compact U(1) gauge fields\cite{Herbut_confinement}.
Condensation of instantons (magnetic
monopoles)\cite{Instanton_monopole} is well known to cause
confinement of charged particles\cite{Polyakov,Fradkin,NaLee},
here Dirac spinons. Recently, Hermele et al. showed that the
algebraic spin liquid can be stable against magnetic monopole
excitations\cite{Hermele_QED3}. Ignoring the compactness of U(1)
gauge fields $a_{\mu}$, one can show that the $S_{ASL}$ has a
nontrivial charged fixed point in two space and one time
dimensions [$(2+1)D$] in the limit of large
flavors\cite{Hermele_QED3,Kim_QED3,Kleinert_QED3}, identified with
the algebraic spin liquid. Hermele et al. showed that the charged
critical point in the case of noncompact U(1) gauge fields can be
stable against magnetic monopole excitations of compact U(1) gauge
fields in the limit of large flavors\cite{Hermele_QED3}.
Condensation of monopoles can be forbidden at the stable charged
fixed point owing to critical fluctuations of Dirac fermions. The
$S_{ASL}$ in Eq. (1) is a critical field theory at the charged
critical point, where correlation functions exhibit power law
behaviors with anomalous critical exponents resulting from long
range gauge interactions\cite{Kim_QED3,QED3_exponent}. This is the
reason why the state described by the $S_{ASL}$ is called the
algebraic spin liquid. In appendix A we briefly discuss how the
effective quantum electrodynamics in $(2+1)D$ ($QED_3$), the
$S_{ASL}$ in Eq. (1) can be derived from the antiferromagnetic
Heisenberg model on square lattices.

However, we should remember that the criticality of algebraic spin
liquid holds only for large flavors of critical Dirac spinons. If
the flavor number is not sufficiently large, the internal charge
$e$ is not screened out satisfactorily by critical Dirac fermions.
Then, gauge interactions can make bound states of Dirac fermions,
resulting in massive Dirac spinons. This is known to be
spontaneous chiral symmetry breaking
($S\chi{S}B$)\cite{CSB1,CSB2,DonKim_QED3,Herbut,Dmitri}. In the
case of physical SU(2) antiferromagnets it is not clear if the
algebraic spin liquid criticality remains owing to the $S\chi{S}B$
causing antiferromagnetism. It is believed that there exists the
critical flavor number $N_c$ associated with $S\chi{S}B$ in the
$QED_3$\cite{CSB1,CSB2}. But, the precise value of the critical
number is far from consensus\cite{CSB2}. If the critical value is
larger than $2$, the $S\chi{S}B$ is expected to occur for the
physical $N = 2$ case. Then, the Dirac fermions become massive. On
the other hand, in the case of $N_c < 2$ the algebraic spin liquid
criticality remains stable against the $S\chi{S}B$.
Experimentally, antiferromagnetic long range order is clearly
observed in the SU(2) antiferromagnet. This leads us to consider
the $S\chi{S}B$ in the algebraic spin liquid. In Eq. (1) $S_{M}$
represents the contribution of a fermion mass due to the
$S\chi{S}B$. $m_{\psi}$ is a mass parameter corresponding to
staggered magnetization in the context of
antiferromagnetism\cite{DonKim_QED3}. ${\vec n}$ represents
fluctuations of Neel order parameter fields regarded as Goldstone
bosons in the $S\chi{S}B$. ${\vec \tau}$ is Pauli matrix acting on
the spin (flavor) space of Dirac spinons. The mass parameter
$m_{\psi}$ can be determined by a self-consistent gap equation,
given by $m_{\psi} \approx e^{2}exp[-2\pi/\sqrt{N_c/N - 1}]$ in
the $1/N$ approximation\cite{CSB1}. For completeness of this paper
we briefly sketch the derivation of dynamical mass generation in
appendix B. There are additional fermion bilinears connected with
the Neel state by "chiral" transformations because the $S_{ASL}$
in Eq. (1) has more symmetries than those of the Heisenberg
model\cite{Herbut,Tanaka_sigma,Hermele_sigma}. These order
parameters are associated with valance bond
orders\cite{Herbut,Tanaka_sigma,Hermele_sigma}. But, in the
present paper we consider only the Neel order parameter in order
to obtain the O(3) nonlinear $\sigma$ model.

Contributions of high energy spinons in $S_{ASL} + S_{M}$ lead to
the $S_{NL\sigma{M}}$ in the gradient expansion\cite{Dmitri,
Tanaka_sigma}. This effective action is nothing but the O(3)
nonlinear $\sigma$ model describing quantum antiferromagnetism. In
the $S_{NL\sigma{M}}$ we introduced a Lagrange multiplier field
$\lambda$ to impose the rigid rotor constraint $|{\vec n}|^2 = 1$.
In appendix C we give a detailed derivation of the nonlinear
$\sigma$ model. The total effective action $S_{ASL} + S_{M} +
S_{NL\sigma{M}}$ called fermionic nonlinear $\sigma$ model in Eq.
(1) naturally describes the quantum antiferromagnetism resulting
from the algebraic spin liquid via $S\chi{S}B$ at half filling.
Based on this fermionic O(3) nonlinear $\sigma$ model we discuss
physics of the quantum antiferromagnetism.

In Eq. (1) we focus our attention on the Kondo-like spin coupling
term $\vec{n}\cdot\bar{\psi}\vec{\tau}\psi$ between
antiferromagnetic spin fluctuations ${\vec n}$ and Dirac fermions
$\psi_{n}$. This term shows that an antiferromagnetic excitation
of spin $1$ can fractionalize into two fermionic spinons of spin
$1/2$. But, long range gauge interactions prohibit
antiferromagnetic spin fluctuations from decaying into spinons.
More quantitatively, massive Dirac spinons should be confined to
form spin $1$ antiferromagnetic fluctuations owing to the effect
of magnetic monopoles of compact U(1) gauge fields $a_{\mu}$.
$S\chi{S}B$ makes the criticality of algebraic spin liquid
disappear, causing massive spinons. These spinons can generate
only the Maxwell kinetic energy for the gauge field $a_{\mu}$ via
particle-hole polarizations. It is well known that this Maxwell
gauge theory shows confinement of charged matter fields in
$(2+1)D$ owing to the condensation of magnetic
monopoles\cite{Polyakov,Fradkin,NaLee}. Spin $1/2$ massive Dirac
spinons are confined to make spin $1$ antiferromagnetic
fluctuations, i.e., ${\vec n} = \langle\bar{\psi}{\vec
\tau}\psi\rangle$, where $\langle...\rangle$ denotes a vacuum
expectation value. A resulting effective field theory for this
antiferromagnet is obtained to be \bqa && Z_{AF} = \int{D {\vec
\pi}} e^{- S_{\pi}} , \nn && S_{\pi} = \int{d^3x}
\frac{Nm_{\psi}}{4\pi}\Bigl( |\partial_{\mu}{\vec \pi}|^{2} +
\frac{({\vec \pi}\cdot\partial_{\mu}{\vec \pi})^{2}}{1 - |{\vec
\pi}|^{2}} \Bigr) \nn && \approx \int{d^3x}
\frac{Nm_{\psi}}{4\pi}\Bigl( |\partial_{\mu}{\vec \pi}|^{2} +
({\vec \pi}\cdot\partial_{\mu}{\vec \pi})^{2} \Bigr) , \eqa where
the Neel vector is given by ${\vec n} = ({\vec \pi}, n^{3})$ with
$n^{3} = \sqrt{1 - |{\vec \pi}|^2}$. In the last line we obtained
an effective field theory for small fluctuations of ${\vec \pi}$
fields around the Neel axis $n^{3}$ in the antiferromagnet. The
${\vec \pi}$ fields are nothing but aniferromagnons, considered to
be spinon-antispinon composites $\pi^{\pm} =
\langle\bar{\psi}\tau^{\pm}\psi\rangle = n_{1} \pm in_{2}$ with
relativistic spectrum, $\omega = k$ in the low energy limit. The
last term in the last line represents interactions between
antiferromagnons. Low energy physics in this conventional quantum
antiferromagnet is well described by the interacting
antiferromagnons\cite{Chakravarty}. This is the result of
conventional antiferromagnetism when we do not consider fermion
zero modes. In the present paper we show that the presence of
fermion zero modes can alter this effective field theory Eq. (2)
{\it at high energies}. We will see that the fermion zero mode can
appear from the Kondo-like coupling term, resulting in magnon
decaying into spinons.

\subsection{Existence of Fermion Zero Modes}

Now we show that {\it a fermion zero mode can arise in a
monopole-skyrmion composite potential}. In order to introduce
monopole potentials we separate the compact U(1) gauge field
$a_{\mu}$ into $a_{\mu} = a_{\mu}^{cl} + a_{\mu}^{qu}$, where
$a_{\mu}^{cl}$ represent magnetic monopole (instanton) potentials
and $a_{\mu}^{qu}$, gaussian quantum fluctuations. In addition, we
consider skyrmion configurations ${\vec n}^{sky}$. The single
monopole and skyrmion potentials are given by $a_{\mu}^{cl} =
a(r)\epsilon_{3\alpha\mu}x_{\alpha}$ and $n_{\mu}^{sky} =
\Phi(r)x_{\mu}$, respectively, where $a(r)$ and $\Phi(r)$ are
proportional to $r^{-2}$ in $r \rightarrow \infty$ with $r =
\sqrt{\tau^{2} + x^2 + y^2}$\cite{Jackiw,Potential}. Integrating
over the Dirac spinon fields in the fermionic nonlinear $\sigma$
model Eq. (1), we obtain the following fermion determinant in the
monopole-skyrmion composite potential, $S_{\psi} = -
Nlndet\Bigl[\gamma_{\mu}(\partial_{\mu} - ia^{cl}_{\mu} -
ia_{\mu}^{qu}) + m_{\psi}{\vec n}^{sky}\cdot{\vec \tau} \Bigr]$.
For the time being, we ignore gaussian quantum fluctuations
$a_{\mu}^{qu}$ because our objective is to find a fermion zero
mode in the single monopole potential $a_{\mu}^{cl}$. In order to
calculate the determinant we solve an equation of motion in the
monopole-skyrmion potential \bqa &&
(\gamma_{\mu}\partial_{\mu}\delta_{nm} +
ia(r)(\gamma\times{x})_{3} +
m_{\psi}\Phi(r){x}_{\mu}\tau^{\mu}_{nm})\psi_{m} = E\psi_{n} . \nn
\eqa Remember that in the absence of the skyrmion potential
($m_{\psi} = 0$), i.e., in the algebraic spin liquid the fermion
zero mode was not found\cite{Marston}. As mentioned in the
introduction, the presence of the Neel vector makes the Dirac
equation (3) have the essentially same structure as that utilized
in the context of high energy
physics\cite{Jackiw,tHooft,Callias,Witten}. SU(2) gauge theory in
terms of massless Dirac fermions and adjoint Higgs fields
interacting via SU(2) gauge fields has been intensively studied in
the presence of Kondo-like isospin couplings between these matter
fields\cite{Jackiw,tHooft,Callias,Witten}. When the Higgs fields
are condensed, topologically nontrivial stable excitations called
't Hooft-Polyakov monopoles can appear\cite{Huang_book,Potential}.
The 't Hooft-Polyakov monopoles consist of gauge-Higgs composite
potentials\cite{Huang_book,Potential}. Jackiw and Rebbi showed
that the Dirac equation in the presence of the Kondo-like isospin
couplings has a fermion zero mode in a 't Hooft-Polyakov monopole
potential\cite{Jackiw}. Here, the Neel order parameter fields play
the same role as the adjoint Higgs fields.

Following Jackiw and Rebbi\cite{Jackiw}, we explicitly demonstrate
that Eq. (3) has a zero mode. We rewrite Eq. (3) in terms of the
two component spinors $\chi^{\pm}_{n}$ with $E = 0$ \bqa &&
(\sigma_{3}\partial_{\tau})_{ij}\chi^{\pm}_{jn} +
(\sigma_{2}\partial_{x})_{ij}\chi^{\pm}_{jn} +
(\sigma_{1}\partial_{y})_{ij}\chi^{\pm}_{jn} \nn && +
iay(\sigma_{2})_{ij}\chi^{\pm}_{jn} -
iax(\sigma_{1})_{ij}\chi^{\pm}_{jn} \pm
m_{\psi}\Phi{x}_{\mu}\chi^{\pm}_{jm}(\tau^{\mu{T}})_{mn}  = 0 .
\nn \eqa Inserting $\chi^{\pm}_{in} =
\mathcal{M}^{\pm}_{im}\tau^{2}_{mn}$ with a two-by-two matrix
$\mathcal{M}^{\pm}$ into the above, we obtain \bqa &&
\sigma_{3}\partial_{\tau}\mathcal{M}^{\pm} +
\sigma_{2}\partial_{x}\mathcal{M}^{\pm} +
\sigma_{1}\partial_{y}\mathcal{M}^{\pm}  \nn && +
iay\sigma_{2}\mathcal{M}^{\pm} - iax\sigma_{1}\mathcal{M}^{\pm}
\mp m_{\psi}\Phi\mathcal{M}^{\pm}{x}_{\mu}\sigma^{\mu} = 0 . \eqa
Now the spin matrices $\tau^{k}$ and the Dirac matrices
$\sigma^{k}$ are indistinguishable\cite{Jackiw}. Finally,
representing the matrix $\mathcal{M}^{\pm}$ in
$\mathcal{M}^{\pm}_{im} = g^{\pm}\delta_{im} +
g_{\mu}^{\pm}\sigma^{\mu}_{im}$, we obtain the coupled equations
of motion for the numbers $g^{\pm}$ and $g_{\mu}^{\pm}$ \bqa &&
(\partial_{\tau} \mp m_{\psi}\Phi\tau){g}^{\pm} - i(\partial_{x} +
iay \pm m_{\psi}\Phi{y})g^{\pm}_{1} \nn && + i(\partial_{y} - iax
\pm m_{\psi}\Phi{x} )g^{\pm}_{2} = 0 , \nn && (\partial_{x} + iay
\mp m_{\psi}\Phi{y})g^{\pm} + i(\partial_{\tau} \pm
m_{\psi}\Phi\tau)g^{\pm}_{1} \nn && - i(\partial_{y} - iax \pm
m_{\psi}\Phi{x} )g^{\pm}_{3} = 0 , \nn && (\partial_{y} - iax \mp
m_{\psi}\Phi{x})g^{\pm} - i(\partial_{\tau} \pm
m_{\psi}\Phi\tau)g^{\pm}_{2} \nn && + i(\partial_{x} + iay \pm
m_{\psi}\Phi{y})g^{\pm}_{3} = 0 , \nn && (\partial_{\tau} \mp
m_{\psi}\Phi\tau)g^{\pm}_{3} + (\partial_{x} + iay \mp
m_{\psi}\Phi{y} )g^{\pm}_{2} \nn && + (\partial_{y} - iax \mp
m_{\psi}\Phi{x} )g^{\pm}_{1} = 0 . \eqa These equations yield the
following zero mode equations \bqa && (\partial_{\tau} +
m_{\psi}\Phi\tau)g^{-} = 0 , \nn && (\partial_{x} + iay +
m_{\psi}\Phi{y})g^{-} = 0 , \nn && (\partial_{y} - iax +
m_{\psi}\Phi{x})g^{-} = 0 . \eqa A normalizable zero mode solution
$g^{-}$ is given by $g^{-} \sim exp\Bigl[-i\int{dx}a(r)y + i
\int{dy}a(r)x \Bigr] exp\Bigl[-\int{d\tau}m_{\psi}\Phi(r)\tau -
\int{dx}m_{\psi}\Phi(r)y - \int{dy}m_{\psi}\Phi(r)x \Bigr]$.
Existence of the zero mode makes the fermion determinant zero in
the single monopole excitation. As a result single monopole
excitations are suppressed by the zero mode and thus, the
condensation of magnetic monopoles is forbidden. The suppression
of monopole condensation results in deconfinement of internally
charged particles, here the Dirac spinons. This deconfined
antiferromagnetism differs from the usual confined one described
by Eq. (2). {\it Antiferromagnetic spin fluctuations fractionalize
into spinons because the U(1) gauge interactions are not confining
any more}. Below we discuss an effective field theory to describe
this unusual antiferromagnetism.

\subsection{Effect of Fermion Zero Modes: 't Hooft Effective Interaction and Deconfinement of Massive Spinons}

In high energy physics it is well known that {\it the contribution
of instantons (monopoles)}\cite{Instanton_monopole} {\it in the
presence of a fermion zero mode gives rise to an effective
interaction to fermions}\cite{tHooft,Witten,Dmitri}. This
interaction is usually called 't Hooft effective interaction. In
order to obtain the effective fermion interaction it is necessary
to average the partition function in Eq. (1) over various
instanton and anti-instanton configurations. Following Ref.
\cite{Dmitri}, we first consider a partition function in a single
instanton potential \bqa && Z_{\psi} =
\int{D\psi_{n}}e^{-\int{d^3x}
\bar{\psi}_{n}\gamma_{\mu}\partial_{\mu}\psi_{n}}\Bigl(m -
V^{I}[\psi_{n}]\Bigr) , \nn && V^{I}[\psi_{n}] = \int{d^3x}\Bigl(
\bar{\psi}_{n}(x)\gamma_{\mu}\partial_{\mu}\Phi^{I}_{n}(x)\Bigr)\nn&&\times
\int{d^3y}\Bigl(\bar{\Phi}_{n}^{I}(y)\gamma_{\mu}\partial_{\mu}\psi_{n}(y)\Bigr)
. \eqa Here $\Phi_{n}^{I}$ is the zero mode obtained from Eq. (7).
A fermion mass $m$ is introduced. Later the chiral limit $m
\rightarrow 0$ will be taken. The effective action including the
effective potential $V^{I}[\psi_{n}]$ in Eq. (8) gives a correct
green function in a single instanton potential\cite{Dmitri},
$S^{I}(x,y) = <\psi_{n}(x)\bar{\psi}_{n}(y)> = -
\frac{\Phi_{n}^{I}(x)\bar\Phi_{n}^{I}(y)}{m} + S_{0}(x,y)$ with
the bare propagator $S_{0}(x,y) =
(\gamma_{\mu}\partial_{\mu})^{-1}\delta(x-y)$. Thus the partition
function in Eq. (8) can be used for average of
instantons\cite{Dmitri}. The partition function in the presence of
$N_{+}$ instantons and $N_{-}$ anti-instantons can be easily built
up\cite{Dmitri} \bqa && Z_{\psi} =
\int{D\psi_{n}}{Da^{qu}_{\mu}}e^{-\int{d^3x}\bar{\psi}_{n}\gamma_{\mu}(\partial_{\mu}
- ia^{qu}_{\mu})\psi_{n}} \nn && \times\Bigl(m -
<V^{I}[\psi_{n}]>\Bigr)^{N_+}\Bigl(m - <V^{I}[\psi_{n}]>
\Bigr)^{N_-} .\eqa Here we admit the noncompact U(1) gauge field
$a^{qu}_{\mu}$ representing gaussian quantum fluctuations. In the
following the index $qu$ is omitted. $\langle...\rangle$ means
averaging over individual instantons. Introducing instanton
averaged nonlocal fermion vertices $Y_{\pm} = - V<V^{I}[\psi_{n}]>
=-\int{d^3z_{I(\bar{I})}}V^{I(\bar{I})}[\psi_{n}]$ with volume
$V$, where $z_{I(\bar{I})}$ represent instanton (anti-instanton)
positions\cite{Dmitri}, we obtain the following partition function
in the chiral limit $m \rightarrow 0$ \bqa && Z_{\psi} =
\int{D\psi_{n}}{Da_{\mu}}
e^{-\int{d^3x}\bar{\psi}_{n}\gamma_{\mu}(\partial_{\mu} -
ia_{\mu})\psi_{n}}\nn&&
\int\frac{d\lambda_{\pm}}{2\pi}\int{d\Gamma_{\pm}}
e^{i\lambda_{+}(Y_{+} - \Gamma_{+}) + N_{+}\ln\frac{\Gamma_{+}}{V}
+ ( + \rightarrow - )} . \eqa  Integration over $\lambda_{\pm}$
and $\Gamma_{\pm}$ recovers Eq. (9) in the chiral limit. In the
thermodynamic limit $N_{\pm}, V \rightarrow \infty$ and
${N_{\pm}}/{V}$ fixed, integration over $\Gamma_{\pm}$ and
$\lambda_{\pm}$ can be performed by the saddle point
method\cite{Dmitri}. Integrating over $\Gamma_{\pm}$ first, we
obtain \bqa && Z_{\psi} =
\int\frac{d\lambda_{\pm}}{2\pi}e^{N_{+}\Bigl(\ln\frac{N_{+}}{i\lambda_{+}V}
- 1 \Bigr) + (+\rightarrow -)}\nn&&
\int{D\psi_{n}}{Da_{\mu}}e^{-\int{d^3x}\bar{\psi}_{n}\gamma_{\mu}(\partial_{\mu}
- ia_{\mu})\psi_{n} + i\lambda_{+}Y_{+} + i\lambda_{-}Y_{-}} .
\eqa An explicit calculation for the instanton average shows that
the vertex $Y^{\pm}$ corresponds to a mass\cite{Dmitri}, $Y^{\pm}
= \int\frac{d^3k}{(2\pi)^{3}}
(2\pi\rho{F}(k))^{2}\bar{\psi}_{n}\frac{1\pm\gamma_{5}}{2}\psi_{n}$
with $\gamma_{5} = \left( \begin{array}{cc} 0 & I \\ -I & 0
\end{array} \right)$. Here $F(k)$ is associated with the fermion
zero mode in the effective potential $V^{I}[\psi_{n}]$ in Eq. (8).
In the present paper we do not perform an explicit calculation for
the instanton average in $Y_{\pm}$ and thus, we do not know the
exact form of $F(k)$. Our objective is to see how the 't Hooft
interaction appears as an instanton effect. Here $\rho$ is the
size of an instanton. Although the instanton (magnetic monopole)
can be considered to be a point particle, the instanton-skyrmion
composite would have its characteristic size $\rho$. In order to
determine the size $\rho$ we should solve an equation of motion
for the Neel vector ${\vec n}$ in the presence of an instanton
configuration. In the present paper we do not examine this issue.
We assume its existence. Owing to the neutrality condition of
magnetic charges, $N_{+} = N_{-} = N_{I}/2$ is obtained in Eq.
(11), where $N_{I}$ is the total number of instantons and
anti-instantons. The saddle point solution of $\lambda_{+} =
\lambda_{-} \equiv \lambda$ in Eq. (11) gives rise to cancellation
of the $\gamma_{5}$ term in the mass, causing a momentum dependent
mass $m(k) = m_{I}F^{2}(k)$ with $m_{I} =
\lambda(2\pi\rho)^{2}$\cite{Dmitri}. The mass $m_{I}$ is
determined by the saddle point equation for $\lambda$ usually
called a self-consistent gap equation\cite{Dmitri} \bqa
&&\frac{8}{N_{I}/V}\int\frac{d^3k}{(2\pi)^{3}}\frac{m^{2}(k)}{k^2+m^2(k)}
= 1 . \eqa Ignoring the momentum dependence by setting $F(k) = 1$
for simplicity, we obtain the 't Hooft mass $m_{I} =
\frac{\pi}{2\Lambda^{1/2}}\Bigl(\frac{N_{I}}{V}\Bigr)^{1/2}$ with
a momentum cut-off $\Lambda$ in the small mass limit. Since the
mean density of instantons is proportional to the instanton
fugacity, $N_{I}/V \sim y_{m} = e^{- S_{inst}}$ with an instanton
action $S_{inst} \sim 1/e^{2}$\cite{Polyakov,NaLee}, the fermion
mass is roughly given by $m_{I} \sim y_{m}^{1/2}$. We can obtain
the following effective Lagrangian in terms of Dirac spinons
$\psi_{n}$ with the 't Hooft effective mass $m_{I}$ interacting
via {\it noncompact} U(1) gauge fields $a_{\mu}$, ${\cal L}_{\psi}
= \bar{\psi}_{n}\gamma_{\mu}(\partial_{\mu} - ia_{\mu})\psi_{n} +
m_{I}\bar{\psi}_{n}\psi_{n}$. The formal appearance of the fermion
mass does not necessarily mean that the fermions are massive. This
is because the fermion mass is determined by density of
instantons. In order to determine the fermion mass, we should find
the instanton fugacity, i.e., reveal the state of instantons.

It is one of the most difficult problems to determine the state of
monopoles (instantons) in the presence of matter fields. In this
paper we do not pursue to determine the state of monopoles
precisely. Instead, we consider all possibilities. Generally
speaking, there would be three possible monopole states. These are
monopole plasma, monopole dipolar and monopole "liquid" phases.
These phases are characterized by $y_{m} \rightarrow \infty$,
$y_{m} \rightarrow 0$ and $y_{m} \not= 0$, respectively, where
$y_{m}$ is the monopole fugacity. We exclude the first, monopole
plasma phase because the presence of fermion zero modes does not
allow the condensation of monopoles.

At first glance a liquid phase of magnetic monopoles sounds quite
strange. It is known that the liquid state does not appear in the
Abelian Higgs model without
fermions\cite{Senthil_deconfinement,Kim1,Kleinert,Ichinose_deconfinement}.
But, in the present case we do not have any evidence to exclude
this monopole state. In $(2+1)D$ the basic trend is confinement,
i.e., $y_m \rightarrow \infty$\cite{Fradkin,NaLee} away from
quantum
criticality\cite{Hermele_QED3,Senthil_deconfinement,Kim1,Kleinert,Ichinose_deconfinement}.
Owing to the confinement tendency many instantons would be excited
although the presence of fermion zero modes does not allow
instanton condensation. Thus, we should treat dense uncondensed
monopoles. We claim that in order to solve this problem a new
methodology beyond the dilute approximation of monopoles is
required. As far as we know, this methodology is not found yet.
Note that the usual renormalization group equation for the
monopole fugacity is obtained from the dilute approximation of
monopoles\cite{Polyakov,Hermele_QED3,Kim_QED3,Kleinert_QED3,Kim1,Kleinert,Kim3}.
It is not clear if this standard renormalization group equation is
applicable to high density limit. In this respect we have no clear
evidence to exclude the monopole liquid phase in the presence of
fermion zero modes. We may view the emergence of the liquid state
as the proximate effect of the Higgs-confinement phase in the
presence of fermion zero modes. There exist some reports about a
new phase instead of plasma and dipolar phases in two dimensional
Coulomb gas when the density of particles is high\cite{KT}.
Furthermore, a new fixed point with nonzero monopole fugacity was
recently reported even in the $QED_3$ only with massless Dirac
fermions\cite{Kleinert_QED3}.

In the liquid phase of monopoles Dirac spinons obtain the 't Hooft
effective mass ($m_{I}$) while in the dipolar phase of monopoles
the 't Hooft mass vanishes. But, it should be noted that we are
considering antiferromagnetism. In antiferromagnetism Dirac
spinons are massive ($m_{\psi}$) via $S\chi{S}B$. This mass
completely differs from the 't Hooft mass. Although it is not easy
to evaluate the 't Hooft mass, we expect that the chiral mass
$m_{\psi}$ would be larger than the 't Hooft mass $m_{I}$. This
assumption is based on the fact that the chiral mass $m_{\psi}$
would be identified with the staggered magnetization observed in
quantum antiferromagnets. In both phases of magnetic monopoles the
deconfined Dirac spinons would be massive owing to the
antiferromagnetism. At present we do not have any idea how to
distinguish these two monopole states.

Combining the existence of fermion zero modes in section II-B and
their physical effects in section II-C, we can reach the following
effective Lagrangian {\it from Eq. (1) in the deconfined
antiferromagnet} \bqa && Z_{AF} = \int{D\psi_{n}}{Da_{\mu}}{D{\vec
\pi}} e^{-\int{d^3x} {\cal L}_{AF}} , \nn && {\cal L}_{AF} =
\bar{\psi}_{n}\gamma_{\mu}(\partial_{\mu} - ia_{\mu})\psi_{n} +
m_{\psi}\bar{\psi}_{n}\tau^{3}_{nm}\psi_{m} +
\frac{1}{2e^2}|\partial\times{a}|^2 \nn && -
\frac{1}{2}m_{\psi}\bar{\psi}_{n}\tau^{3}_{nm}\psi_{m}|{\vec
\pi}|^{2} + m_{\psi}\bar{\psi}_{n}{\vec
\tau}_{nm}\psi_{m}\cdot{\vec \pi} \nn && +
\frac{Nm_{\psi}}{4\pi}\Bigl( |\partial_{\mu}{\vec \pi}|^{2} +
({\vec \pi}\cdot\partial_{\mu}{\vec \pi})^{2} \Bigr) , \eqa where
we used $n^{3} = \sqrt{1-|{\vec \pi}|^2} \approx 1 -
\frac{1}{2}|{\vec \pi}|^{2}$ as Eq. (2). In this effective
Lagrangian we should understand that the gauge field $a_{\mu}$ is
{\it noncompact} in Eq. (13) owing to the effect of fermion zero
modes while the gauge field $a_{\mu}$ in Eq. (1) is {\it compact},
causing confinement of Dirac spinons and thus, {\it resulting in
Eq. (2)}. In this antiferromagnet low lying excitations are
antiferromagnetic spin waves and (noncompact) U(1) gauge
fluctuations. Both gapless excitations cause $C_{v} \sim T^{2}$,
where $C_{v}$ is specific heat and $T$,
temperature\cite{DonKim_QED3}. Furthermore, the gapped Dirac
spinons would be deconfined to emerge above their mass gap. These
gapped spinons interact with the antiferromagnons as shown by the
fourth and fifth terms\cite{LsigmaM}. In Fig. 1 we show schematic
spin spectrum in inelastic neutron scattering. Here
${\chi"}(\omega)$ is the imaginary part of the transverse dynamic
spin susceptibility at momentum $(\pi,\pi)$, given by
$\chi"(\omega) \approx \chi"_{\pi}(\omega) + \chi"_{\psi}(\omega)
+ \chi"_{c}(\omega)$ where $\chi"_{\pi} \sim
\langle\pi^{+}\pi^{-}\rangle$ is a magnon susceptibility,
$\chi"_{\psi} \sim \langle{\bar \psi}\tau^{+}\psi{\bar
\psi}\tau^{-}\psi\rangle$, a spinon susceptibility, and $\chi"_{c}
\sim \langle{\bar \psi}\tau^{+}\psi\pi^{-}\rangle +
\langle\pi^{+}{\bar \psi}\tau^{-}\psi\rangle$, a "coupling"
susceptibility in a highly schematic form. {\it At low energies
antiferromagnons give dominant spectral weight ($\chi"_{\pi}$)
while at high energies Dirac spinons exhibit broad continuum
spectrum ($\chi"_{\psi}$)}. In our scenario the unexplained $1/2$
spectral weight in the inelastic neutron
scattering\cite{Experiment} would be identified with the gapped
Dirac spinons. Although scattering between spinons and magnons in
Eq. (13) does not alter the spin spectrum in Fig. 1 {\it
qualitatively}, elaborate calculations are required in order to
understand the spin spectrum more quantitatively and precisely.

\begin{figure}
\includegraphics[width=8cm]{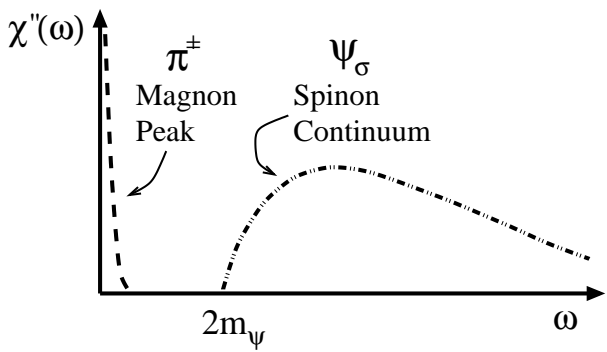}
\caption{\label{Fig. 1} Schematic spin spectrum in inelastic
neutron scattering }
\end{figure}

\section{Summary}

In summary, we investigated {\it the two dimensional quantum
antiferromagnet resulting from the algebraic spin liquid via
spontaneous chiral symmetry breaking}. This antiferromagnet is
described by {\it the fermionic nonlinear $\sigma$ model}, Eq. (1)
in terms of Dirac spinons $\psi_{n}$ interacting via not only U(1)
gauge fluctuations $a_{\mu}$ but also antiferromagnetic spin
fluctuations ${\vec n}$. We showed that {\it the Kondo-like spin
couplings} between the Dirac spinons and Neel vectors give rise to
{\it the fermion zero mode} in the single monopole potential. The
existence of fermion zero modes suppresses the condensation of
monopoles, thus causing the deconfinement of spinons. As a result
{\it we obtained the effective field theory Eq. (13) in the
deconfined antiferromagnet, which differs from the conventional
confined antiferromagnet described by the O(3) nonlinear $\sigma$
model Eq. (2)}. From the effective field theory Eq. (13) we argued
that {\it the deconfined massive spinons} would be observed as
{\it particle-hole continuum above their mass gap} in the dynamic
spin susceptibility\cite{Experiment,Fisher,YBKim}.

\section{Acknowledgement}

K. -S. Kim is much indebted to Dr. A. Tanaka who pointed out a
mistake in association with the gradient expansion in Eq. (C3) and
Eq. (C4). K.-S. Kim thanks Dr. Yee, Ho-Ung for correcting English
errors. K.-S. Kim also appreciates helpful discussion with Dr.
Park, Tae-Sun.

\appendix
\section{}

In appendix A we briefly sketch how we can obtain the $S_{ASL}$ in
Eq. (1) from the antiferromagnetic Heisenberg model on two
dimensional square lattices, $H =
J\sum_{<i,j>}\vec{S}_{i}\cdot\vec{S}_{j}$ with $J>0$. Inserting
the following spinon representation for spin, $\vec{S}_{i} =
\frac{1}{2}f^{\dagger}_{i\alpha}\vec{\tau}_{\alpha\beta}f_{i\beta}$
into the above Heisenberg model, and performing the standard
Hubbard-Stratonovich transformation for an exchange interaction
channel, we obtain an effective one body Hamiltonian for fermions
coupled to an order parameter, $H_{eff} =
-J\sum_{<i,j>}f_{i\alpha}^{\dagger}\chi_{ij}f_{j\alpha} - h.c.$.
Here $f_{i\alpha}$ is a fermionic spinon with spin $\alpha =
\uparrow, \downarrow$, and $\chi_{ij}$ is an auxiliary field
called a hopping order parameter. Notice that the hopping order
parameter $\chi_{ij}$ is a complex number defined on links $ij$.
Thus, it can be decomposed into $\chi_{ij} =
|\chi_{ij}|e^{i\theta_{ij}}$, where $|\chi_{ij}|$ and
$\theta_{ij}$ are the amplitude and phase of the hopping order
parameter, respectively. Inserting this representation for the
$\chi_{ij}$ into the effective Hamiltonian, we obtain $H_{eff} =
-J\sum_{<i,j>}|\chi_{ij}|f_{i\alpha}^{\dagger}e^{i\theta_{ij}}f_{j\alpha}
- h.c.$. We can easily see that this effective Hamiltonian has an
internal U(1) gauge symmetry, $H'_{eff}[f'_{i\alpha},\theta'_{ij}]
= H_{eff}[f_{i\alpha},\theta_{ij}]$ under the following U(1) phase
transformations, $f'_{i\alpha} = e^{i\phi_{i}}f_{i\alpha}$ and
$\theta'_{ij} = \theta_{ij} + \phi_{i} - \phi_{j}$. This implies
that the phase field $\theta_{ij}$ of the hopping order parameter
plays the same role as a U(1) gauge field $a_{ij}$. When a spinon
hops on lattices, it obtains an Aharnov-Bohm phase owing to the
U(1) gauge field $a_{ij}$. It is known that a stable mean field
phase is a $\pi$ flux state at half
filling\cite{Review,DonKim_QED3}. This means that a spinon gains
the phase of $\pi$ when it turns around one plaquette. In the
$\pi$ flux phase low energy elementary excitations are massless
Dirac spinons near nodal points showing gapless Dirac spectrum and
U(1) gauge fluctuations\cite{Review,DonKim_QED3}. In the low
energy limit the amplitude $|\chi_{ij}|$ is frozen to $|\chi_{ij}|
= J|<f_{j\alpha}^{\dagger}f_{i\alpha}>|$. A resulting effective
field theory for one possible quantum disordered paramagnetism of
the antiferromagnetic Heisenberg model is $QED_3$ in terms of
massless Dirac spinons interacting via compact U(1) gauge fields,
$S_{ASL}$ in Eq. (1).

In the $S_{ASL}$ $\psi_{n} = \left(
\begin{array}{c} \chi^{+}_{n} \\  \chi^{-}_{n} \end{array} \right)$
is a four component massless Dirac fermion, where $n = 1, 2$
represents its SU(2) spin ($\uparrow, \downarrow$) and $\pm$
denote the nodal points of $(\pi/2,\pm\pi/2)$ in momentum space.
Usually, SU(N) quantum antiferromagnets are considered by
generalizing the spin components $n = 1, 2$ into $n = 1, 2, ...,
N$. The two component spinors $\chi^{\pm}_{n}$ are given by
$\chi^{+}_{1} = \left(
\begin{array}{c} f_{\uparrow{1e}} \\ f_{\uparrow{1o}} \end{array}
\right)$, $\chi^{-}_{1} = \left( \begin{array}{c} f_{\uparrow{2o}}
\\ f_{\uparrow{2e}} \end{array} \right)$,
$\chi^{+}_{2} = \left( \begin{array}{c} f_{\downarrow{1e}}
\\ f_{\downarrow{1o}} \end{array} \right)$, and
$\chi^{-}_{2} = \left( \begin{array}{c} f_{\downarrow{2o}} \\
f_{\downarrow{2e}} \end{array} \right)$, respectively. In the
spinon field $f_{abc}$ $a = \uparrow, \downarrow$ represents its
SU(2) spin, $b = 1, 2$, the nodal points ($+, -$), and $c = e, o$,
even and odd sites, respectively\cite{DonKim_QED3}. Dirac matrices
$\gamma_{\mu}$ are given by $\gamma_{0} = \left( \begin{array}{cc}
\sigma_{3} & 0 \\ 0 & -\sigma_{3} \end{array} \right)$,
$\gamma_{1} = \left( \begin{array}{cc} \sigma_{2} & 0 \\ 0 &
-\sigma_{2} \end{array} \right)$, and $\gamma_{2} = \left(
\begin{array}{cc} \sigma_{1} & 0 \\ 0 & -\sigma_{1} \end{array}
\right)$, respectively, where they satisfy the Clifford algebra
$[\gamma_{\mu},\gamma_{\nu}]_{+} =
2\delta_{\mu\nu}$\cite{DonKim_QED3}.

\section{}

In appendix B, for completeness of this paper we briefly sketch
how we obtain the dynamically generated spinon mass $m_{\psi}$ in
$S_{M}$ in Eq. (1). A single spinon propagator is given by
$G^{-1}(k) = G_{0}^{-1}(k) - \Sigma(k)$, where $G_{0}^{-1}(k) =
i\gamma_{\mu}k_{\mu}$ is the inverse of a bare spinon propagator,
and $\Sigma(k)$, a spinon self-energy resulting from long range
gauge interactions. The spinon self-energy is determined by the
self-consistent gap equation, $\Sigma(k) =
\int\frac{d^3q}{(2\pi)^{3}}Tr[\gamma_{\mu}G(k-q)\gamma_{\nu}D_{\mu\nu}(q)]$,
where $D_{\mu\nu}(q)$ is a renormalized propagator of the U(1)
gauge field $a_{\mu}$ due to particle-hole excitations of massless
Dirac fermions. The self-energy can be written as $\Sigma(k) = -
m_{\psi}(k)\tau^{3}$ for staggered
magnetization\cite{DonKim_QED3}. Inserting this representation
into the above self-consistent gap equation, we obtain the
following expression for the spinon mass, $m_{\psi}(p) =
\int\frac{d^3k}{(2\pi)^{3}}\frac{\gamma_{\mu}m_{\psi}(k)\gamma_{\nu}}{k^{2}
+ m_{\psi}^{2}(k)}D_{\mu\nu}(p-k)$. The renormalized gauge
propagator $D_{\mu\nu}(q)$ is obtained to be $D_{\mu\nu}(q)
\approx \Pi_{\mu\nu}^{-1}(q) = \frac{8}{Nq}\Bigl(\delta_{\mu\nu} -
\frac{q_{\mu}q_{\nu}}{q^2}\Bigr)$ in the Lorentz
gauge\cite{DonKim_QED3}, where $\Pi_{\mu\nu}(q) = -
N\int\frac{d^3k}{(2\pi)^{3}}Tr[\gamma_{\mu}G_{0}(k)\gamma_{\nu}G_{0}(k-q)]$
is the polarization function of massless Dirac fermions in the
$1/N$ approximation. Inserting this gauge propagator into the gap
equation and performing an angular integration, one can find
$m_{\psi}(p) =
\frac{4}{N\pi^2p}\int_{0}^{\Lambda}dk\frac{km_{\psi}(k)}{k^{2} +
m_{\psi}^{2}(k)}(k+p-|k-p|)$, where $\Lambda$ is a momentum
cutoff\cite{DonKim_QED3}. This integral expression is equivalent
to the differential equation,
$\frac{d}{dp}\Bigl(p^{2}\frac{dm_{\psi}(p)}{dp}\Bigr) = -
\frac{8}{\pi^2N}\frac{p^2m_{\psi}(p)}{p^2+m_{\psi}^2(p)}$ with
boundary conditions, $\Lambda\frac{dm_{\psi}(p)}{dp}_{p = \Lambda}
+ m_{\psi}(\Lambda) = 0$ and $0 \leq m_{\psi}(0) <
\infty$\cite{CSB1,DonKim_QED3}. In Ref. \cite{CSB1} this equation
is well analyzed in detail. Its solution is given by $m_{\psi}
\approx e^{2}exp[-2\pi/\sqrt{N_c/N - 1}]$ in the case of $N <
N_c$\cite{CSB1,Herbut}.

\section{}

In appendix C we discuss how the O(3) nonlinear $\sigma$ model can
be derived from the effective action $S_{ASL} + S_{M}$ in Eq. (1).
Integration over the Dirac fermions in $S_{ASL} + S_{M}$ results
in the following expression \bqa && S_{NL\sigma{M}} = -
ln\Bigl[\int{D\psi_n}
exp\{-\int{d^3x}\Bigl(\bar{\psi}_{n}\gamma_{\mu}(\partial_{\mu} -
ia_{\mu})\psi_{n} \nn && +  m_{\psi}\bar{\psi}_{n}({\vec
n}\cdot{\vec \tau}_{nm})\psi_{m} \Bigr)\}\Bigr] \nn && = -
Nln\det\Bigl[\gamma_{\mu}(\partial_{\mu} - ia_{\mu}) +
m_{\psi}{\vec n}\cdot{\vec \tau} \Bigr] \nn && \approx -
\frac{N}{2}ln\det(-\partial^{2} + m_{\psi}^{2}) \nn && +
\int{d^3x}\Bigl( \frac{Nm_{\psi}}{4\pi}|\partial_{\mu}{\vec
n}|^{2} + \frac{N}{12\pi{m}_{\psi}}|\partial\times{a}|^{2} \Bigr)
. \eqa The second term in the last line is well derived in Ref.
\cite{Dmitri}, \bqa && - \frac{N}{2}ln\det\Bigl[1 -
\frac{m_{\psi}\gamma_{\mu}\partial_{\mu}({\vec n}\cdot{\vec
\tau})}{-\partial^{2} + m_{\psi}^{2}}\Bigr] \nn && = -
\frac{N}{2}Tr\int{d^3x}\langle{x}|ln\Bigl[1 -
\frac{m_{\psi}\gamma_{\mu}\partial_{\mu}({\vec n}\cdot{\vec
\tau})}{-\partial^{2} + m_{\psi}^{2}}\Bigr]|x\rangle \nn && = -
\frac{N}{2} \int{d^3x}\int\frac{d^3k}{(2\pi)^{3}} \nn &&
e^{-ikx}Trln\Bigl[1 -
\frac{m_{\psi}\gamma_{\mu}\partial_{\mu}({\vec n}\cdot{\vec
\tau})}{-\partial^{2} + m_{\psi}^{2}}\Bigr]e^{ikx} \nn && = -
\frac{N}{2} \int{d^3x}\int\frac{d^3k}{(2\pi)^{3}} \nn &&
Trln\Bigl[1 - \frac{m_{\psi}\gamma_{\mu}\partial_{\mu}({\vec
n}\cdot{\vec \tau})}{k^{2} + m_{\psi}^{2} -
2ik_{\mu}\partial_{\mu} -
\partial^{2}}\Bigr] \nn && \approx \int{d^3x}
\frac{N}{4}\int\frac{d^3k}{(2\pi)^3}
Tr\Bigl(\frac{m_{\psi}\gamma_{\mu}\partial_{\mu}({\vec
n}\cdot{\vec \tau})}{k^{2} + m_{\psi}^{2}}\Bigr)^{2} \nn && =
\int{d^3x} \frac{Nm_{\psi}}{4\pi}|\partial_{\mu}{\vec n}|^{2} .
\eqa In the above $Tr$ stands for not a functional but a usual
matrix trace for both flavor (spin) and spinor indices. In going
from the third to the fourth line we have dragged the factor
$e^{ikx}$ through the operator, thus shifting all differential
operators $\partial_{\mu} \rightarrow
\partial_{\mu} + ik_{\mu}$\cite{Dmitri}. Expanding the argument of the
logarithmic term in powers of $\partial_{\mu}{\vec n}$ and of
$2ik_{\mu}\partial_{\mu} + \partial^{2}$, one can easily obtain
the expression in the fifth line. Performing the momentum
integration, we obtain an effective spin stiffness proportional to
the mass parameter $m_{\psi}$. This implies that the rigidity of
fluctuations in the Neel field is controlled by the mass parameter
$m_{\psi}$ of Dirac spinons. Eq. (C2) is nothing but the O(3)
nonlinear $\sigma$ model describing quantum antiferromagnetism.
Note that higher order derivative terms in the gradient expansion
are irrelevant in $(2+1)D$ in the renormalization group sense.

Next, we sketch the derivation of the Maxwell gauge action.
Expanding the argument of the logarithmic term in Eq. (C1) to the
second order of the gauge field $a_{\mu}$, we obtain $S_{gauge} =
\int\frac{d^3q}{(2\pi)^3}\frac{1}{2}a_{\mu}(q)\Pi_{\mu\nu}(q)a_{\nu}(-q)$,
where the fermion polarization function $\Pi_{\mu\nu}(q)$ is given
by $\Pi_{\mu\nu}(q) = - N
\int\frac{d^3k}{(2\pi)^3}Tr[\gamma_{\mu}G(k+q)\gamma_{\nu}G(k)]$
with the single spinon propagator $G(k) = [i\gamma_{\mu}k_{\mu} +
m_{\psi}{\vec n}\cdot{\vec \tau}]^{-1}$. Utilizing the Feynman
identity and trace identity of Dirac gamma matrices, one can
obtain the following expression for the polarization function,
$\Pi_{\mu\nu}(q) =
2N(TrI)\frac{\Gamma(2-D/2)}{(4\pi)^{D/2}}(q^{2}\delta_{\mu\nu} -
q_{\mu}q_{\nu})\int_{0}^{1}dx(1-x)x(m_{\psi}^2+q^2x(1-x))^{D/2-2}
= \frac{(TrI)N}{4\pi}(q^{2}\delta_{\mu\nu} -
q_{\mu}q_{\nu})\Bigl(\frac{m_{\psi}}{2q^2} +
\frac{q^2-4m_{\psi}^2}{4q^3}\sin^{-1}\Bigl(\frac{q}{\sqrt{4m_{\psi}^2+q^2}}\Bigr)\Bigr)$\cite{DonKim_QED3}.
This leads to the Maxwell gauge action in Eq. (C1).

Lastly, we should comment the reason why imaginary terms do not
arise in the present nonlinear $\sigma$ model because some
previous studies have shown the emergence of imaginary
terms\cite{Abanov_sigma,Chiral_anomaly}. Following the evaluations
in Ref. \cite{Abanov_sigma}, one can obtain two imaginary terms
{\it in the irreducible representation of gamma matrices}; one is
a coupling term $ia_{\mu}J_{\mu}$ between topologically nontrivial
fermionic currents $J_{\mu} =
\frac{1}{8\pi}\epsilon_{\mu\nu\lambda}
\epsilon_{\alpha\beta\gamma} n^{\alpha}\partial_{\nu}n^{\beta}
\partial_{\lambda}n^{\gamma}$ and U(1) gauge fields $a_{\mu}$, and
the other, a geometrical phase $iN\pi\Gamma[{\vec
n}]$\cite{Abanov_sigma,Chiral_anomaly}. {\it The key point is the
representation of Dirac gamma matrices}\cite{A_Tanaka}. Here, we
utilized four-by-four Dirac matrices by combining the two nodal
points. Note that the signs of Pauli matrices in the Dirac gamma
matrices are opposite for the nodal points $\pm$. This fact
results in cancellation of the imaginary terms. The first
imaginary term can be considered from a variation of the
logarithmic term in Eq. (C1) with respect to the gauge field
$a_{\mu}$ \bqa &&
iNTr\Bigl[\gamma_{\mu}\delta{a}_{\mu}\frac{1}{\gamma_{\mu}(\partial_{\mu}
- ia_{\mu}) + m_{\psi}({\vec n}\cdot{\vec \tau})} \Bigr] \nn && =
iNTr\Bigl[\gamma_{\mu}\delta{a}_{\mu}\frac{-\gamma_{\mu}(\partial_{\mu}
- ia_{\mu}) + m_{\psi}({\vec n}\cdot{\vec
\tau})}{-[\gamma_{\mu}(\partial_{\mu} - ia_{\mu})]^{2} +
m_{\psi}^{2} - m_{\psi}\gamma_{\mu}\partial_{\mu}({\vec
n}\cdot{\vec \tau})} \Bigr] \nn && \approx
iNTr\Bigl[\gamma_{\mu}\delta{a}_{\mu}\frac{m_{\psi}({\vec
n}\cdot{\vec \tau})}{-\partial^{2} +
m_{\psi}^{2}}\Bigl(\frac{m_{\psi}\gamma_{\mu}\partial_{\mu}({\vec
n}\cdot{\vec \tau})}{-\partial^{2} + m_{\psi}^{2}}\Bigr)^{2}\Bigr]
. \eqa In this expression the key point is the triple product of
Dirac matrices, $\gamma_{\mu}\gamma_{\nu}\gamma_{\lambda}$ in the
last line. In the irreducible representation of gamma matrices,
i.e., Pauli matrices, this contribution is nonzero, leading to
$\epsilon_{\mu\nu\lambda}$. As a result the imaginary term of
$i{a}_{\mu}\epsilon_{\mu\nu\lambda} \epsilon_{\alpha\beta\gamma}
n^{\alpha}\partial_{\nu}n^{\beta}
\partial_{\lambda}n^{\gamma}$ can
be obtained\cite{Abanov_sigma,Chiral_anomaly}. Another
$\epsilon_{\alpha\beta\gamma}$ associated with the Neel vectors
appears from the triple product of Pauli matrices,
$\tau_{\alpha}\tau_{\beta}\tau_{\gamma}$. On the other hand, in
the present representation of Dirac matrices the contribution of
the $+$ nodal point leads to $+\epsilon_{\mu\nu\lambda}$ while
that of the $-$ nodal point, $-\epsilon_{\mu\nu\lambda}$. Thus,
{\it these two contributions are exactly cancelled}. The
geometrical phase term, considered from a variation of the
logarithmic term in Eq. (C1) with respect to the Neel field ${\vec
n}$\cite{Abanov_sigma,Chiral_anomaly}, \bqa && -
NTr\Bigl[m_{\psi}\delta({\vec n}\cdot{\vec
\tau})\frac{1}{\gamma_{\mu}\partial_{\mu} + m_{\psi}({\vec
n}\cdot{\vec \tau})} \Bigr] \nn && = -
NImTr\Bigl[m_{\psi}\delta({\vec n}\cdot{\vec
\tau})\frac{-\gamma_{\mu}\partial_{\mu} + m_{\psi}({\vec
n}\cdot{\vec \tau})}{-\partial^{2} + m_{\psi}^{2} -
m_{\psi}\gamma_{\mu}\partial_{\mu}({\vec n}\cdot{\vec \tau})}
\Bigr] \nn && \approx NTr\Bigl[m_{\psi}\delta({\vec n}\cdot{\vec
\tau})\frac{m_{\psi}({\vec n}\cdot{\vec \tau})}{-\partial^{2} +
m_{\psi}^{2}}\Bigl(\frac{m_{\psi}\gamma_{\mu}\partial_{\mu}({\vec
n}\cdot{\vec \tau})}{-\partial^{2} + m_{\psi}^{2}}\Bigr)^{3}\Bigr]
, \eqa is also exactly zero owing to the same reason. {\it Another
way to say this is that the signs of mass terms for the Dirac
fermions ($\chi_{n}^{+}$ and $\chi_{n}^{-}$) at the two Dirac
nodes ($+$ and $-$) are opposite, resulting in cancellation of the
parity anomaly}\cite{A_Tanaka}. If we fix the Neel vector in the
$z$ direction (${\vec n} = {\hat z}$), we can see {\it the
opposite signs} explicitly from $m_{\psi}
\bar{\psi}_{n}\tau^{z}_{nm}\psi_{m} = m_{\psi}
{\psi}^{\dagger}_{n}\gamma_{0}\tau^{z}_{nm}\psi_{m} =
m_{\psi}\chi^{+}_{n}\sigma_{z}\tau^{z}_{nm}\chi^{+}_{m} -
m_{\psi}\chi^{-}_{n}\sigma_{z}\tau^{z}_{nm}\chi^{-}_{m} =
m_{\psi}{\bar\chi}^{+}_{n}\tau^{z}_{nm}\chi^{+}_{m} -
m_{\psi}{\bar\chi}^{-}_{n}\tau^{z}_{nm}\chi^{-}_{m}$. Both massive
Dirac fermions ($\chi_{n}^{\pm}$) contribute to the imaginary
terms, respectively. However, the signs of the imaginary terms are
opposite and thus, the cancellation occurs. As a result the
imaginary terms do not appear in the present nonlinear $\sigma$
model. This was already discussed in Ref. \cite{Zou,Wen}.

It is possible that the mass terms have the same signs.
Considering the two gamma matrices of $\gamma_{4} = \left(
\begin{array}{cc} 0 & I \\ I & 0 \end{array} \right)$ and $\gamma_{5} = \left(
\begin{array}{cc} 0 & I \\ - I & 0 \end{array} \right)$\cite{DonKim_QED3}, we can obtain
the following mass terms with the same signs, ${\cal
\tilde{L}}_{M} =
m_{\psi}\bar{\psi}_{n}\gamma_{4}\gamma_{5}\tau^{z}\psi_{m} = -
m_{\psi}{\bar\chi}^{+}_{n}\tau^{z}_{nm}\chi^{+}_{m} -
m_{\psi}{\bar\chi}^{-}_{n}\tau^{z}_{nm}\chi^{-}_{m}$. These mass
terms can arise from the algebraic spin liquid, $S_{ASL}$ in Eq.
(1) via $S\chi{S}B$ because the algebraic spin liquid has the
enlarged symmetry\cite{Tanaka_sigma,Hermele_sigma}, as discussed
earlier. In this case the cancellation dose not occur and thus,
the imaginary terms necessarily arise. This antiferromagnetism
would not be conventional since it breaks not only time reversal
symmetry but also parity symmetry. When this antiferromagnetism
disappears via strong quantum fluctuations, its corresponding
quantum disordered paramagnet is expected to be the {\it chiral
spin liquid}\cite{Tanaka_sigma,Zou,Wen}. In the present paper we
did not discuss the chiral spin liquid.


\begin{thebibliography}{9}
\bibitem{Instanton_monopole} In the present paper we use the
terminology "magnetic monopole" instead of "instanton" in
association with confinement. "Instanton" is utilized in
association with 't Hooft effective interaction.
\bibitem{LsigmaM} These two terms are known to
be $\pi\pi{N}{N}$ and $\pi{NN}$ scattering, respectively, in the
context of high energy physics, where $\pi$ is a meson field and
$N$, a nucleon field.
\end{thebibliography}
\end{document}